\documentclass[runningheads]{llncs}

\usepackage[utf8]{inputenc}
\usepackage[T1]{fontenc}
\usepackage{amsfonts}
\usepackage{amsmath}
\usepackage{hyperref}
\usepackage[capitalise]{cleveref}
\usepackage{placeins}
\usepackage{float}
\usepackage{cprotect}

\usepackage{cite} 
\newcommand{\citep}{\cite} 

\usepackage{graphicx}
\usepackage{booktabs}
\usepackage{caption}        
\usepackage{subcaption}     
\usepackage{multirow}
\usepackage{xcolor}
\usepackage{nicefrac,xfrac}
\usepackage{microtype}
\usepackage{contour}
\usepackage{adjustbox}

\begin{document}
\title{Domain-specific augmentations with resolution agnostic self-attention mechanism improves choroid segmentation in optical coherence tomography images}
\titlerunning{REACHNet: Robust, Efficient \& self-Attentive CHoroid analysis in OCT}

\makeatletter
\newcommand{\printfnsymbol}[1]{%
  \textsuperscript{\@fnsymbol{#1}}%
}
\makeatother

\author{Jamie Burke\thanks{Corresponding and lead author.}\inst{1,2}, 
Justin Engelmann\inst{3,4}, 
Charlene Hamid\inst{5}, 
Diana Moukaddem\inst{6},
Dan Pugh\inst{7},
Neeraj Dhaun\inst{7},
Amos Storkey\inst{3},
Niall Strang\inst{6},
Stuart King\inst{1}, 
Tom MacGillivray\inst{5,8}, 
Miguel O. Bernabeu\inst{3,9} \and
Ian J.C. MacCormick\inst{8,10}}

\authorrunning{J. Burke et al.}

\institute{
School of Mathematics, University of Edinburgh, Edinburgh, UK \\
\email{jamie.burke@ed.ac.uk}\and
Robert O Curle Ophthalmology Suite, Institute for Regeneration and Repair, University of Edinburgh, UK \and
School of Informatics, University of Edinburgh, Edinburgh, UK \and
Centre for Medical Informatics, University of Edinburgh, Edinburgh, UK \and 
Clinical Research Facility and Imaging, University of Edinburgh, Edinburgh, UK\and 
Department of Vision Sciences, Glasgow Caledonian University, Glasgow, UK\and 
British Heart Foundation Centre for Cardiovascular Science, University of Edinburgh, Edinburgh, UK\and
Centre for Clinical Brain Sciences, University of Edinburgh, Edinburgh, UK \and
The Bayes Centre, University of Edinburgh, Edinburgh, UK \and 
Institute for Adaptive and Neural Computation, School of Informatics, University of Edinburgh, Edinburgh, UK
}

\maketitle

\begin{abstract}
The choroid is a key vascular layer of the eye, supplying oxygen to the retinal photoreceptors. Non-invasive enhanced depth imaging optical coherence tomography (EDI-OCT) has recently improved access and visualisation of the choroid, making it an exciting frontier for discovering novel vascular biomarkers in ophthalmology and wider systemic health. However, current methods to measure the choroid often require use of multiple, independent semi-automatic and deep learning-based algorithms which are not made open-source. Previously, Choroidalyzer --- an open-source, fully automatic deep learning method trained on 5,600 OCT B-scans from 385 eyes --- was developed to fully segment and quantify the choroid in EDI-OCT images, thus addressing these issues. Using the same dataset, we propose a \textbf{R}obust, \textbf{R}esolution-agnostic and \textbf{E}fficient \textbf{A}ttention-based network for \textbf{CH}oroid segmentation (\textbf{REACH}). REACHNet leverages multi-resolution training with domain-specific data augmentation to promote generalisation, and uses a lightweight architecture with resolution-agnostic self-attention which is not only faster than Choroidalyzer's previous network (4 images/s vs. 2.75 images/s on a standard laptop CPU), but has greater performance for segmenting the choroid region, vessels and fovea (Dice coefficient for region 0.9769 vs. 0.9749, vessels 0.8612 vs. 0.8192 and fovea 0.8243 vs. 0.3783) due to its improved hyperparameter configuration and model training pipeline. REACHNet can be used with Choroidalyzer as a drop-in replacement for the original model and will be made available upon publication.

\end{abstract}

\section{Introduction}

The choroid, a dense vascular layer behind the retina, plays a pivotal role in the maintenance of the outer retinal layers \cite{nickla2010multifunctional}. Non-invasive optical coherence tomography (OCT) images the choroid at micron resolution, but its optical signal has historically struggled to penetrate this tissue due to the hyperreflective retinal-pigment epithelium (RPE) sitting above it, thus leaving OCT imaging previously focused on retinal image analysis. Recently, enhanced depth imaging OCT (EDI-OCT) has improved choroidal visualisation \cite{spaide2008enhanced}, enabling researchers to investigate the choroid's potential as a biomarker in systemic health research and broader `oculomics' \cite{robbins2021choroidal, burke2023evaluation, tan2016state}.

The choroid contains a dense, heterogeneous population of vessels interspersed and suspended by connective tissue in its interstitial space. Isolating the choroidal region and vessels are the two primary tasks in OCT choroid analysis. Segmenting the region relies on detecting a single shape using identifiable landmarks, such as its position below the hyperreflective RPE and above the homogeneous appearing sclera. Conversely, defining choroidal vessels is far more challenging due to speckle noise and the complex nature of the vascular bed, meaning OCT images capture oblique sections through vessels which often don't have clear boundaries. This can make manual segmentation prohibitively time consuming, and potentially lead to inconsistencies among existing segmentation methods \cite{wei2018comparison}.

Accurate choroid analysis demands precise segmentation of the region and vessels, and careful consideration of the region of interest used for analysis. This typically involves referencing the fovea's location and accounting for image quality variations. This standardisation is crucial for consistent comparisons across diverse studies, populations and imaging devices. Unfortunately, current practices often rely on separate tools for full segmentation and measurement of the choroid \cite{liu2019robust, burke2021edge, khaing2021choroidnet, muller2022application, burke2023evaluation, burke2023opensource} and fovea localisation \cite{xuan2023deep}. Moreover, many of these tools are not openly available to researchers which can further hinder accessibility and standardisation in the field \cite{mazzaferri2017open, kugelman2019automatic, chen2022application}.

\begin{figure}[tb]
    \centering
    \includegraphics[width=\textwidth]{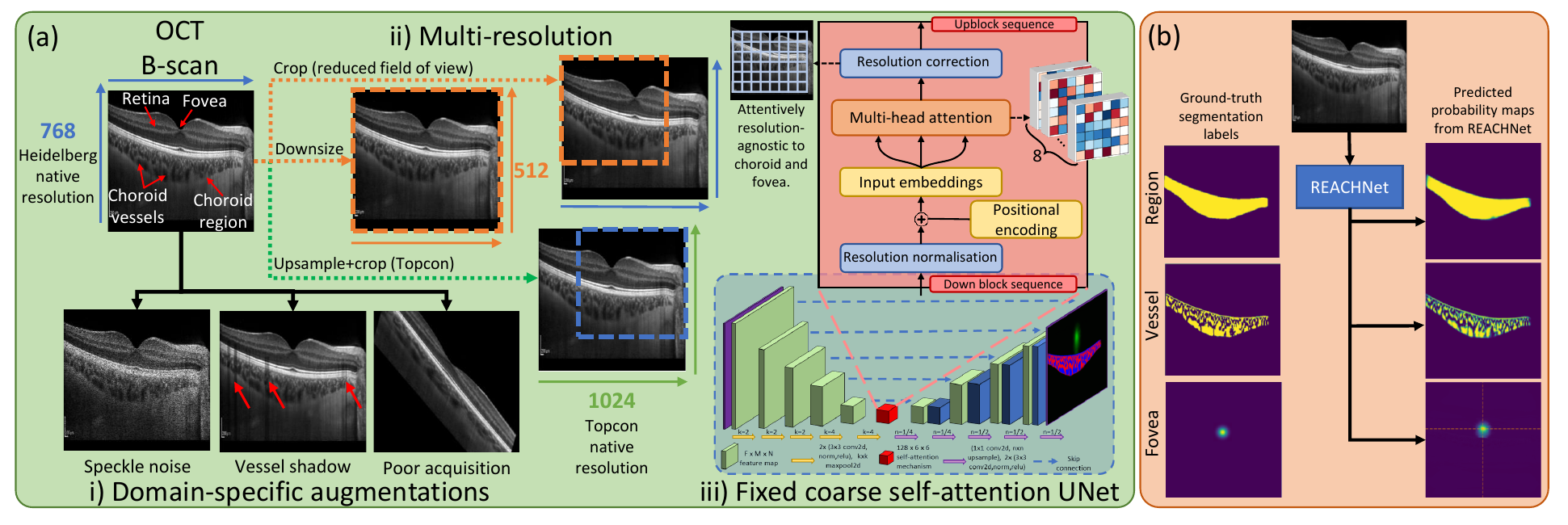}
    \cprotect\caption{Summary of the improvements we propose to build REACHNet (a), including domain-specific data augmentation (i), multi-resolution training (ii) and a resolution-agnostic self-attention mechanism at the deepest point in REACHNet's architecture (iii). Ground-truth label segmentations, and the output expected by REACHNet (b).}
    \label{fig:infographic}
\end{figure}

Choroidalyzer \cite{engelmann2023choroidalyzer} is a recent deep learning approach to segment and measure the choroidal region, vessels, and fovea in OCT images. We use REACHNet to improve on Choroidalyzer's model on the same dataset (\cref{fig:infographic}). REACHNet utilises multi-resolution training and domain-specific augmentations to generalise across device resolutions, handle speckle artefacts, vessel occlusions and poor image acquisition. Its lightweight architecture with a resolution-agnostic self-attention mechanism improves segmentation performance compared to Choroidalyzer and other pre-trained models, while reducing inference speed and memory consumption through an optimised training pipeline and hyperparameter configuration.

\section{Methods}

\subsection{Data}

The protocol for data collection is previously described \cite{engelmann2023choroidalyzer}. Briefly, we used 5,600 OCT B-scans belonging to 385 eyes of 233 subjects using 5 cohorts related to systemic health (chronic kidney disease \cite{dhaun2014optical}, dementia \cite{ritchie2012prevent}, pregnancy and diurnal variation \cite{moukaddem2022comparison}). B-scans were collected from three imaging devices from two leading OCT imaging device manufacturers, Heidelberg Engineering ---- the Standard Heidelberg spectral domain OCT SPECTRALIS Module and its mobile-equivalent, the SPECTRALIS FLEX Module (Heidelberg Engineering, Heidelberg, Germany) --- and Topcon, the swept source OCT DRI Triton Plus (Topcon, Tokyo, Japan). Our dataset comprises macular-centred, horizontal line B-scans from Heidelberg Engineering imaging devices, and macular, fovea-centred radial B-scans from the Topcon DRI Triton Plus device. Supplementary Table 1 provides an overview of the population used in this work. 

5 of the 6 cohorts were split at the patient level into training (4,144 B-scans, 122 subjects), validation (466 B-scans, 28 subjects) and internal test sets (756 B-scans, 37 subjects). The OCTANE cohort was entirely held-out and used as an external test set (168 B-scans, 46 subjects). Supplementary Table 2 describes the populations within each set. 

We follow the same protocol which Choroidalyzer uses for generating ground-truth choroid region segmentation labels \cite{engelmann2023choroidalyzer, burke2023opensource}. Choroidal vessel segmentation labels were generated with a software tool previously described in \cite{engelmann2023choroidalyzer}, after the following pre-processing steps: images were enhanced using non-local means denoising, contrast limited histogram equalisation and Gamma-level transforms to sharpen choroidal vessel walls, and a column-wise pixel intensity moving average to compensate for superficial retinal vessel shadowing was applied. The fovea pixel coordinate was manually selected as in \cite{engelmann2023choroidalyzer}, but here each segmentation mask is an isotropic 2D Gaussian centred at this coordinate with a standard deviation of 30 pixels --- B-scans not fovea-centred set the origin as the corresponding mask.

\subsection{Domain-specific data augmentation}

\textbf{Speckle noise.} OCT B-scans suffer from multiplicative speckle noise from scattering light waves emitted from the device during acquisition. We recreate this by degrading a B-scan using element-wise multiplication with an array whose elements are independently sampled from a Gaussian distribution $\mathcal{N}(0, \mathcal{U}(0.2, 0.5))$.

\textbf{Vessel shadowing.} A primary problem of choroid vessel segmentation is where incident light perpendicularly penetrates through superficial retinal vessels, subsequently darkening the corresponding A-scans (pixel columns). We artificially produce these shadows through randomly selecting $n$ A-scans uniformly and lower the brightness of a random number $w$ of adjacent A-scans using element-wise Gamma transforms ($n, w \sim \mathcal{U}(2, 10)$). The beginning of each shadow is 30 pixels above the choroid, approximately at the outer plexiform layer which are where the A-scan columns are first corrupted from retinal vessels.

\textbf{Acquisition distortion.} Optimal OCT image acquisition expects the retinal and choroidal structures to lie parallel to the horizontal image axis. However, various factors including patient concentration, imager inexperience and degree of myopathy can skew the B-scan. We use an affine transformation with some random degree of rotation $\theta \in \mathcal{U}(5,45)$ and anisotropic scaling $(s_x, s_y)$ to simulate this acquisition, where $s_x, s_y$ are the image axes pixel length scales. The ratio $s_x : s_y$ is approximately $4 : 1$ for Heidelberg and Topcon devices, so the vertical and horizontal image axes will be squashed and elongated as the image is rotated.

\subsection{Multi-resolution training} 
The native resolutions for the Heidelberg Standard and FLEX Modules are $768 \times 768$ and $496 \times 768$, respectively, while the Topcon DRI Triton Plus has a native resolution of $992 \times 1024$. Choroidalyzer's experiments were conducted and evaluated at a common $768 \times 768$ pixel resolution \cite{engelmann2023choroidalyzer}, leading to unexpected behaviour when applied at higher resolution (Topcon) or to a reduced field of view. Here, we either fix the input resolution of each batch to $768 \times 768$ or to one of $1024 \times 1024$ or $512 \times 512$ with equal probability ($P=\nicefrac{1}{3})$. For the larger resolution, we crop a random $768 \times 768$ patch to reduce computational load, and for the smaller resolution we either resize or randomly crop the image to $512 \times 512$ ($P=\nicefrac{1}{2}$), the latter option simulating a reduced field of view.

\subsection{Model architecture with resolution-agnostic attention mechanism}

The UNet architecture \cite{ronneberger2015u} of Choroidalyzer's deep learning model has been described previously \cite{engelmann2023choroidalyzer}. While the block structure is kept fixed, REACHNet uses a reduced depth size of 5 and scales the feature maps of the first 3 blocks by 2 and the remaining blocks by 4. This reduces the feature map of the deepest block to $6 \times 6$ (for an input resolution of $768 \times 768$) to enable global contextual learning. To account for the reduced depth, our filter size doubles every block from 8 until a maximum of 128, utilising skip connections between symmetrical down- and up-blocks to aid information flow and enhance representation power. 

REACHNet has a multi-head, self-attention block \cite{vaswani2017attention} with 8 heads at the deepest point of the architecture. The self-attention mechanism is resolution agnostic and is applied at a fixed resolution of $6 \times 6$, resizing feature maps accordingly in order to handle multi-resolution input during training. We hypothesise that contextually weighting local patches in the deepest feature maps would aid memory consumption, multi-resolution segmentation and fovea localisation.

\subsection{Model training and hyperparameter configuration}

All our experiments used the following training pipeline: Each is trained for 40 epochs using the AdamW optimizer \cite{loshchilov2017decoupled} (weight decay $1\times10^{-8}$, $\beta_1=0.9, \beta_2=0.999$) minimising the binary cross entropy loss. We used a linear warm-up of the learning rate from $5\times10^{-4}$ to $5\times10^{-3}$ for 5 epochs, and then decayed it following a cosine relationship for 25 epochs, after which it was fixed at $5\times10^{-4}$ for 10 epochs. We used a batch size of 8 to lower memory consumption but used an effective batch size of 64 by linearly scaling the learning rate by $\nicefrac{1}{8}$ \cite{goyal2017accurate}. To smooth output segmentations, we maintained an exponential moving average of the model weights from epoch 20 using a decay of 0.99. Images were standardised using a mean and standard deviation of 0.5. Alongside our domain-specific augmentations (all $P=\nicefrac{1}{3}$), we applied horizontal flipping ($P=\nicefrac{1}{2}$), random perspective shifts with a scale of 0.2 ($P=\nicefrac{1}{4}$), altered the brightness and contrast randomly in [0.6, 1.4] ($P=\nicefrac{1}{3}$) and applied isotropic scaling in [0.9, 1.5], rotation in [-25, 25] and shear in [-15, 15] ($P=\nicefrac{1}{3}$). Affine transforms were selected instead of our acquisition distortion with equal probability ($P=\nicefrac{1}{2}$).  We used python 3.12, PyTorch 2.0, Segmentation Models PyTorch \cite{Iakubovskii2019} and the timm library \cite{rw2019timm}.

\subsection{Evaluation and statistical analysis}
We compare REACHNet and our proposed improvements against Choroidalyzer's underlying model, as well as two ImageNet-pretrained UNet models using ResNet-18 \cite{he2016deep} and MobileNetV3 \cite{howard2019searching} backbones. The dice coefficient was used to measure segmentation performance, and clinically relevant, fovea-centred choroid-derived measurements of thickness, area and vascular index were assessed using Pearson correlation coefficients and mean absolute error (MAE). We follow the same protocol which Choroidalyzer uses for measuring choroid thickness, area and vascular index \cite{engelmann2023choroidalyzer, burke2023evaluation}. All metrics for both internal and external test sets were evaluated on images at their native resolution.

As vessel segmentation is the most challenging task, we had a clinical ophthalmologist qualitatively assess segmentations from REACHNet against Choroidalyzer \cite{engelmann2023choroidalyzer} and the ground-truth labels, in which it was trained, in a masked a randomised fashion. For 100 examples with the poorest agreement, they graded the quality of the image and identified which segmentation they preferred. They also qualitatively rated the segmentations of 50 images with the greatest disagreement, according to their sensitivity (vessel detection) and specificity (interstitial exclusion), using an ordinal rating system from -2 (very bad) to 2 (very good).

\section{Results}

\textbf{Speed and memory consumption.} All experiments were run on a computer with a 13$^\text{th}$ generation, Intel Z1 processor running at 2GHz on Windows 10 with 32Gb of RAM, utilising an NVIDIA GeForce RTX 3070 graphics card with 8Gb of VRAM. We used dummy data with a resolution of $768 \times 768$ to measure each model's performance (\cref{tab:perf_tab}). REACHNet is significantly more compute efficient for training and inference, irrespective of processing unit. Moreover, on a standard laptop CPU, REACHNet can process an image in 0.256 $\pm$ 0.004s, a significant improvement on Choroidalyzer's model, taking 0.362 $\pm$ 0.490s.

\begin{table*}[tb]
\centering
{\small
\scalebox{0.75}{
\centerline{\begin{tabular}{@{}llllllll@{}}
\toprule
  \multirow{2}{2cm}{Model} & 
  \multirow{2}{1.25cm}{Disk size (Mb)} &
  \multicolumn{3}{c}{Inference} &
  \multicolumn{2}{c}{Training} \\
  \cmidrule(l){3-5}\cmidrule(l){6-7}
  \multicolumn{1}{c}{} &
  \multicolumn{1}{c}{} &
  \multicolumn{1}{c}{CPU (img/s)} &
  \multicolumn{1}{c}{GPU (img/s)} &
  \multicolumn{1}{c}{Memory (Gb)} &
  \multicolumn{1}{c}{GPU (time/epoch)}&
  \multicolumn{1}{c}{Memory (Gb)}\\
   \midrule
Choroidalyzer \cite{engelmann2023choroidalyzer} &  3.196 & 14.54 & 252.14 & 1.61 & 3m 26.72s  & \textbf{2.39} \\
UNet w/ ResNet18 \cite{Iakubovskii2019}  & 56.082 & 4.84& 303.28& 2.05 & 3m 29.09s& 3.79\\
UNet w/ MobileNetV3 \cite{Iakubovskii2019} & 4.160 & 12.77& 94.42& \textbf{1.41} & 3m 58.11s& 2.93\\
REACHNet \textrm{(ours)} & \textbf{3.191}  & \textbf{19.41} & \textbf{326.25} & 1.60  & \textbf{2m 42.50s} & 2.76 \\
\bottomrule
\end{tabular}}
}
}
\caption{Speed and memory consumption of the models used in all experiments.}
\label{tab:perf_tab}
\end{table*}

\textbf{Agreement in segmentation.} REACHNet out-performed Choroidalyzer \cite{engelmann2023choroidalyzer} across all segmentation tasks. This can especially be observed for the fovea (internal test 0.8379 vs. 0.3788, external 0.8243 vs. 0.3783) and vessels (internal 0.8445 vs. 0.8236, external 0.8612 vs. 0.8192), with region detection reaching similar, but higher, levels of performance (internal 0.9792 vs. 0.9753, external 0.9769 vs. 0.9749). Importantly, REACHNet had excellent performance on the external test set, whose cohort and disease type share no overlap to the training data, even against pre-trained UNet models across most metrics.

\begin{table*}[tb]
\centering
{\small
\scalebox{0.8}{\centerline{\begin{tabular}{@{}llllllll@{}}
\toprule
  \multirow{3}{2cm}{Experiment} &
  \multicolumn{3}{c}{Internal} &
  \multirow{1}{0.25cm}{} &
  \multicolumn{3}{c}{External}\\
  \cmidrule(l){2-4}\cmidrule(l){6-8}
  \multicolumn{1}{c}{} &
  \multicolumn{1}{c}{Region} &
  \multicolumn{1}{c}{Vessel} &
  \multicolumn{1}{c}{Fovea} &
  \multirow{1}{0.25cm}{} &
  \multicolumn{1}{c}{Region} &
  \multicolumn{1}{c}{Vessel} &
  \multicolumn{1}{c}{Fovea} \\ 
  \midrule
Choroidalyzer \cite{engelmann2023choroidalyzer} & 0.9753 & 0.8236  & 0.3788 & &  0.9749& 0.8192 & 0.3783 \\
UNet w/ ResNet18 \cite{Iakubovskii2019}  & 0.9763  & 0.8404  & 0.8066 &  & 0.9763  & 0.8504 & 0.8066 \\
UNet w/ MobileNetV3 \cite{Iakubovskii2019}   & 0.9499  & 0.7881  & 0.8303 & & 0.9499     & 0.7881  & \textbf{0.8303}\\
REACHNet \textrm{(ours)} & \textbf{0.9792}  & \textbf{0.8445} & \textbf{0.8379} &  & \textbf{0.9769}   & \textbf{0.8612}   & 0.8243 \\
\midrule
\multicolumn{1}{l}{REACHNet --- no domain-specific augs} & 0.9773  & \underline{0.8467} & 0.8273 &    & 0.9742  & 0.8587   & 0.8065 \\
\multicolumn{1}{l}{REACHNet --- no multi-resolution} & 0.9774 & 0.8445 & 0.8281 & & 0.975 & 0.8571 & \underline{0.8488}\\
\multicolumn{1}{l}{REACHNet ---  no self-attention}  & 0.9790  & 0.8456  & 0.8146 &  & 0.9747 & 0.8553 & 0.8193\\
\bottomrule
\end{tabular}}
}
}
\caption{Segmentation performance using Dice coefficient in the internal and external test sets.}
\label{tab:segperf_tab}
\end{table*}

\textbf{Agreement in choroid-derived measurements.} REACHNet had stronger Pearson correlation coefficients with all clinically meaningful, segmentation-derived choroid measurements of interest compared with Choroidalyzer \cite{engelmann2023choroidalyzer} and previously state of the art pre-trained UNet models (\cref{tab:chorperf_tab}). While we observed significantly lower correlations for choroid vascular index (CVI) compared with other metrics in both test sets (internal test Pearson area, 0.9852; thickness, 0.9838; fovea, 0.9912; CVI, 0.6146), REACHNet obtained better mean absolute errors, CA: 0.035mm$^2$ vs. 0.0417mm$^2$; CT: 8.36$\mu$m vs. 9.70$\mu$m; CVI: 0.038 vs. 0.0767; fovea $xy$: (5.0px, 5.5px) vs. (5.9px, 12.6px)) (see supplementary Table 3 for an ablated table of segmentation-derived MAE values). Importantly, the values reported here are impressively small compared to repeatability measurements previously reported in the literature \cite{ma2024validation, rahman2011repeatability}. 

\begin{table*}[tb]
\centering
{\small
\scalebox{0.8}{
\centerline{
\begin{tabular}{@{}llllllllll@{}}
\toprule
\multicolumn{1}{c}{\multirow{2}{*}{Experiment}} & \multicolumn{4}{c}{Internal} &   \multirow{1}{0.5cm}{}
 & \multicolumn{4}{c}{External} \\
\cmidrule(l){2-5}\cmidrule(l){7-10}
\multicolumn{1}{c}{} & \multicolumn{1}{c}{Area} & \multicolumn{1}{c}{Thickness} & \multicolumn{1}{c}{CVI} & \multicolumn{1}{c}{Fovea}  & \multirow{1}{0.5cm}{}
 & \multicolumn{1}{c}{Area} & \multicolumn{1}{c}{Thickness} & \multicolumn{1}{c}{CVI} & \multicolumn{1}{c}{Fovea} \\
\midrule
Choroidalyzer \cite{engelmann2023choroidalyzer} & 0.9728 & 0.9716 & 0.4162 & 0.986 &  & 0.9751 & 0.9773 & 0.5976 & 0.8383 \\
UNet w/ ResNet18 \cite{Iakubovskii2019} & 0.9835 & 0.9819 & 0.5953 & 0.9621 &  & 0.9735 & 0.9819 & 0.6149 & 0.8728 \\
UNet w/ MobileNetV3 \cite{Iakubovskii2019} & 0.7609 & 0.7573 & 0.3041 & 0.9712 &  & 0.7609 & 0.7573 & 0.3041 & 0.8896 \\
REACHNet \textrm{(ours)} & \textbf{0.9852} & \textbf{0.9836} & \textbf{0.6146} & \textbf{0.9912} &  & \textbf{0.9810} & \textbf{0.9828} & \textbf{0.6281} & \textbf{0.8751} \\
\midrule
\multicolumn{1}{l}{REACHNet --- no domain-specific augs} & 0.985 & 0.9832 & \underline{0.6186} & 0.9912 &  & 0.9717 & 0.9752 & 0.5906 & 0.8795 \\
\multicolumn{1}{l}{REACHNet --- no multi-resolution} & 0.9472 & 0.9450 & 0.5765 & 0.9163 &  & 0.9736 & 0.9774 & 0.5345 & \underline{0.8798} \\
\multicolumn{1}{l}{REACHNet --- no self-attention} & 0.9841 & \underline{0.9838} & 0.5812 & 0.9778 &  & 0.9759 & 0.9782 & 0.5327 & 0.8804 \\
\bottomrule
\end{tabular}}
}}
\caption{Pearson correlation of choroid measurements between experiments and ground-truth labels in both test sets. CVI, choroid vascular index.}
\label{tab:chorperf_tab} 
\end{table*}

\textbf{Qualitative evaluation.} REACHNet had an 85\% preference rating which was invariant to image quality (\cref{tab:quality_tab}) and device type (supplementary Table 4). Qualitative grading of the 50 segmentations with greatest disagreement implies the models' improved sensitivity and specificity over Choroidalyzer \cite{engelmann2023choroidalyzer}, being primarily rated ``Good'' for vessel detection and interstitial exclusion (41 and 42, respectively), and without any negative ratings unlike Choroidalyzer and the ground-truth labels (19 and 20 ``Bad'' ratings in total, respectively).

\begin{table*}[tb]
\centering
{\small

\scalebox{0.7}{\centerline{\begin{tabular}{@{}lccc@{}}
\toprule
Image Quality & REACHNet \textrm{(ours)} & Choroidalyzer \cite{engelmann2023choroidalyzer} & Ground-truth labels \\
\midrule
Very good ($N=2$) & \textbf{2} & 0 & 0\\ 
Good ($N=20$) & \textbf{16} & 2 & 2 \\ 
Okay ($N=29$) & \textbf{24} & 0 & 5\\ 
Bad ($N=36$) & \textbf{31} & 3 & 2 \\ 
Very bad ($N=13$) & \textbf{12} & 1 & 0\\ 
\midrule
Total ($N=100$) & \textbf{85} & 6 & 9\\
\midrule
& & & \\
\midrule
Method & Choroid vessel detection & & Interstitial exclusion\\ \midrule
\multicolumn{1}{l}{\textbf{REACHNet} \textrm{(ours)}} & \textbf{VG: 0, G: 41, O: 9, B: 0, VB: 0} & & \textbf{VG: 1, G: 42, O: 7, B: 0, VB: 0} \\
\multicolumn{1}{l}{Choroidalyzer \cite{engelmann2023choroidalyzer}} & VG: 0, G: 10, O: 25, B: 15, VB: 0 & & VG: 0, G: 31, O: 15, B: 4, VB: 0 \\
\multicolumn{1}{l}{Ground-truth labels} & VG: 0, G: 29, O: 17, B: 4, VB: 0 & & VG: 0, G: 19, O: 15, B: 16, VB: 0 \\ 
\bottomrule
\end{tabular}}}

}
\cprotect\caption{Preference of clinical ophthalmologist for 100 examples with poorest vessel segmentation agreement between REACHNet, Choroidalyzer \cite{engelmann2023choroidalyzer} and ground-truth labels, stratified by image quality (top). Qualitative grading of the 50 vessel segmentations with greatest disagreement for each method (bottom).}
\label{tab:quality_tab} 
\end{table*}

\begin{figure}[tb]
    \centering
    \includegraphics[width=0.925\textwidth]{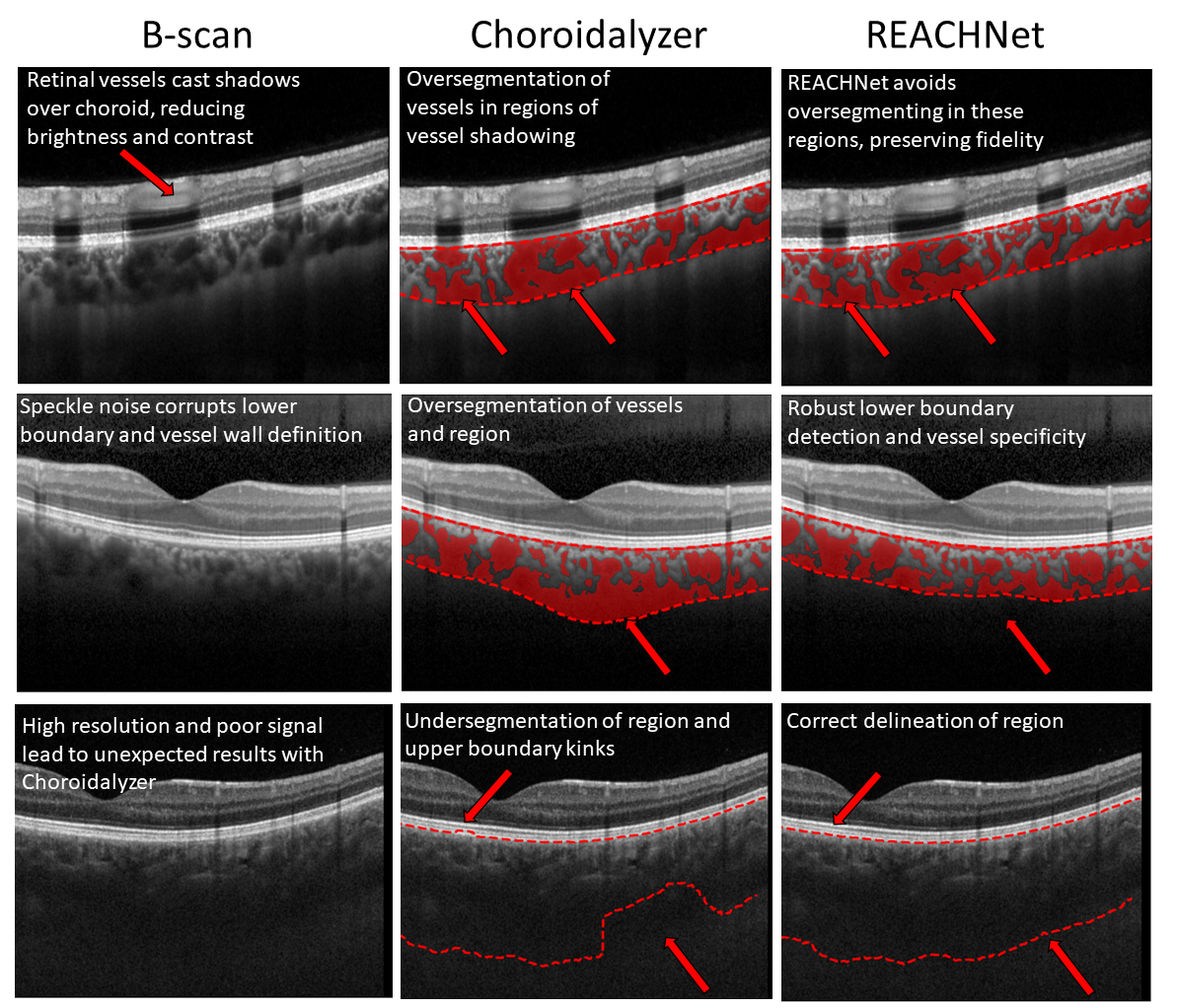}
    \caption{Domain-specific augmentations improve region and vessel segmentation performance of REACHNet over Choroidalyzer's underlying model.}
    \label{fig:augs_examples}
\end{figure}

\section{Discussion}

We have developed REACHNet to deliver a more robust and efficient model for fully automatic segmentation of the choroid in OCT B-scans. We anticipate that REACHNet's model can be used to improve Choroidalyzer's ability to convert a raw OCT B-scan into a set of clinically meaningful segmentation-derived measurements in real-time, thus addressing the distinct lack of unified and open-source methods for choroid analysis.

REACHNet is more robust to artefacts and obscurations commonly observed in OCT B-scans than previously before, producing higher quality segmentations capable of overcoming speckle noise, poor image acquisition and superficial retinal vessel shadowing (\cref{fig:augs_examples}). Moreover, REACHNet can handle multi-resolution input from both Heidelberg and Topcon devices more reliably than before (supplementary Table 4). We therefore expect it to be robust to images of varying quality and resolution from a range of cohorts related to systemic health. 

Interestingly, excluding domain-specific augmentations improved vessel segmentation performance against the ground-truth labels (\cref{tab:segperf_tab}). We believe this --- as well as obtaining only moderately strong correlations of CVI (\cref{tab:chorperf_tab}) --- is due to real vessel shadowing still corrupting the ground-truth segmentation labels. This is evidenced by the ground-truth labels receiving a large number of ``Okay'' and ``Bad'' ratings (15 and 16, respectively) for its specificity (interstitial exclusion). Thus, as REACHNet was overwhelmingly preferred by the clinical ophthalmologist, we believe the superior model of choice is one which includes our domain-specific augmentations, which yielded only positive ratings for the models' interstitial exclusion (specificity), and vessel detection (sensitivity).

REACHNet can process approximately 4 images/s using a standard laptop CPU, compared to Choroidalyzer's 2.75 images/s. This will have major benefits for large-scale data analysis such as the UK Biobank or AlzEye \cite{wagner2022alzeye}. In the latter, their 1,567,358 OCT B-scans would take almost two days less time to process using REACHNet (111 hours) compared to the original model (157 hours), with GPU acceleration significantly reducing this processing time even further.

REACHNet was trained and evaluated using systemic health data. In future, we will address its performance against ocular disease, particularly in the context of choroid-related disease causing extreme choroidal thinning and thickening, such as pathological myopia and central serous chorioretinopathy, respectively. 

REACHNet provides a unified approach to choroid segmentation in OCT, and is freely available to use as Choroidalyzer's deep learning model for researchers and clinicians alike. We hope to bring consistency and robustness, both at scale, to this nascent field.

\section*{Acknowledgements}
J.B. was supported by the Medical Research Council (grant MR/N013166/1) as part of the Doctoral Training Programme in Precision Medicine at the Usher Institute, University of Edinburgh. For the purpose of open access, the authors have applied a creative commons attribution (CC BY) licence to any author accepted manuscript version arising. The authors would also like to thank all participants in the studies used in this paper, as well as all staff at the Edinburgh Imaging Facility who contributed to image acquisition for this study.

\bibliographystyle{splncs04}
\bibliography{references}
\vfill

\setcounter{figure}{0}
\renewcommand{\thefigure}{S\arabic{figure}}
\setcounter{table}{0}
\renewcommand{\thetable}{S\arabic{table}}
\pagebreak

\section{Supplementary materials}

\subsection{Population tables}

\begin{table}[H]
\vspace*{-\baselineskip}
\centering
{\small
\scalebox{0.8}{\centerline{\begin{tabular}{@{}lcccccc|c@{}}
\toprule
 & OCTANE \cite{dhaun2014optical} & Diurnal Variation \cite{dhaun2014optical} & Normative & i-Test \cite{dhaun2014optical} & Prevent Dementia \cite{ritchie2012prevent} & GCU Topcon \cite{moukaddem2022comparison} & Total \\
  \midrule
\multicolumn{1}{l}{Subjects} & 46 & 20 & 1 & 21 & 121 & 24 & 233\\
\multicolumn{1}{r}{Control/Case} & 0 / 46 & 20 / 0 & 1 / 0 & 11 / 10 & 56 / 65 & 24 / 0 & 112 / 121 \\
\multicolumn{1}{r}{Male/Female} & 24 / 22 & 11 / 9 & 1 / 0 & 0 / 21 & 66 / 55 & 14 / 9 & 116 / 116 \\
\multicolumn{1}{r}{Right/Left eyes} & 46 / 0 & 20 / 0 & 1 / 1 & 21 / 21 & 117 / 115 & 22 / 21 & 227 / 158 \\
\multicolumn{1}{r}{Age (mean (SD))} & 47.5 (12.3) & 21.4 (2.3) & 23.0 (0.0) & 32.8 (5.4) & 50.8 (5.6) & 21.8 (7.9) & 42.9 (13.7) \\
\multicolumn{1}{l}{Device manufacturer} & Heidelberg & Heidelberg & Heidelberg & Heidelberg & Heidelberg & Topcon & All \\
\multicolumn{1}{l}{Device type} & Standard & Standard & FLEX & FLEX & Standard & DRI Triton Plus & All \\
\multicolumn{1}{l}{nEDI / EDI} & EDI & EDI & Both & EDI & Both &  &  \\
\multicolumn{1}{l}{Average ART} & 100 & 100 & 9 & 50 & 100 &  &  \\
\multicolumn{1}{l}{Scan location} &  &  &  &  & & & \\
\multicolumn{1}{r}{Horizontal/Vertical} & 168 / 0 & 55 / 50 & 4 / 4 & 76 / 76 & 381 / 369 & 132 / 139 & 816 / 638 \\
\multicolumn{1}{r}{Volume/Radial/Peripapillary} & 0 / 0 / 0 & 0 / 0 / 66 & 365 / 0 / 0 & 2,408 / 0 / 0 & 0 / 0 / 0 & 0 / 1,307 / 0 & 2,773 / 1,307 / 66 \\
\multicolumn{1}{r}{Total B-scans} & 168 & 171 & 373 & 2,560 & 750 & 1,578 & 5,600\\
\bottomrule
\end{tabular}}}}
\caption{Overview of population characteristics. SD, standard deviation. Note that one participant's sex from the GCU Topcon cohort was not recorded. ART, automatic real time B-scan averaging; EDI, enhanced depth imaging. EDI and ART values are not applicable to the Topcon device.}
\label{tab:demo_tab}
\end{table}

\begin{table}[H]
\vspace*{-\baselineskip}
\centering
{\small
\scalebox{0.8}{\centerline{\begin{tabular}{@{}lcccc|c@{}}
\toprule
 & Training & Validation & Testing & External test & Total \\
 \midrule
\multicolumn{1}{l}{Subjects} & 122 & 28 & 37 & 46 & 233 \\
\multicolumn{1}{r}{Male/Female} & 64 / 57 & 12 / 16 & 16 / 21 & 24 / 22 & 116 / 116 \\
\multicolumn{1}{r}{Control/Case} & 76 / 46 & 16 / 12 & 20 / 17 & 0 / 46 & 112 / 121 \\
\multicolumn{1}{r}{Right/Left eyes} & 117 / 107 & 27 / 23 & 37 / 28 & 46 / 0 & 227 / 158 \\
\multicolumn{1}{r}{Standard/FLEX/DRI Triton Plus} & 88 / 14 / 20 & 24 / 2 / 2 & 29 / 6 / 2 & 46 / 0 / 0 & 187 / 22 / 24 \\
\multicolumn{1}{r}{Heidelberg/Topcon} & 102 / 20 & 26 / 2 & 35 / 2 & 46 / 0 & 209 / 24 \\
\multicolumn{1}{r}{Age (mean (SD))} & 40.7 (14.2) & 42.5 (11.9) & 44.5 (13.4) & 47.5 (12.3) & 42.9 (13.4) \\
\multicolumn{1}{l}{Cohort} &  &  &  &  &  \\
\multicolumn{1}{r}{OCTANE} & 0 & 0 & 0 & 46 & 46 \\
\multicolumn{1}{r}{Diurnal Variation} & 12 & 4 & 4 & 0 & 20 \\
\multicolumn{1}{r}{Normative} & 1 & 0 & 0 & 0 & 1 \\
\multicolumn{1}{r}{i-Test} & 13 & 2 & 6 & 0 & 21 \\
\multicolumn{1}{r}{Prevent Dementia} & 76 & 20 & 25 & 0 & 121 \\
\multicolumn{1}{r}{GCU Topcon} & 20 & 2 & 2 & 0 & 24 \\
\multicolumn{1}{l}{B-scans} &  &  &  &  &  \\
\multicolumn{1}{r}{Standard/Flex/DRI Triton Plus} & 582 / 2,281 / 1,281 & 136 / 190 / 140 & 137 / 462 / 157 & 168 / 0 / 0 & 1,023 / 2,933 / 1,578 \\
\multicolumn{1}{r}{Heidelberg/TopCon} & 2,863 / 1,281 & 326 / 140 & 599 / 157 & 168 / 0 & 3,956 / 1,578 \\
&  &  &  &  & \\
\multicolumn{1}{r}{Horizontal/Vertical scans} & 462 / 461 & 90 / 82 & 95 / 95 & 168 / 0 & 816 / 638 \\
\multicolumn{1}{r}{Volume/Radial/Peripapilary scans} & 2,161 / 1,060 / 39  & 178 / 116 / 15 & 434 / 131 / 12 & 0 / 0 / 0 & 2,773 1,307 / 0\\
\multicolumn{1}{r}{Total B-scans} & 4,183 & 481 & 768 & 168 & 5,600 \\
\bottomrule
\end{tabular}}}
}
\caption{Overview of population and image characteristics of the internal training, validation and test sets, and external test set. Note that one participant's sex from the Topcon cohort was not recorded.}
\label{supp_tab:trainvaltest_population_overview_tab}
\end{table}

\subsection{Mean absolute error analysis of choroid measurements}

\begin{table}[H]
\centering
\vspace*{-\baselineskip}
{\small
\scalebox{0.8}{
\centerline{
\begin{tabular}{@{}llllllllll@{}}
\toprule
\multicolumn{1}{c}{Experiment} & \multicolumn{4}{c}{Internal} & \multicolumn{1}{c}{} & \multicolumn{4}{c}{External}\\
\cmidrule(l){2-5}\cmidrule(l){7-10}
\multicolumn{1}{c}{} & \multicolumn{1}{c}{Area} & \multicolumn{1}{c}{Thickness} & \multicolumn{1}{c}{CVI} & \multicolumn{1}{c}{Fovea}  & \multirow{1}{0.5cm}{}
 & \multicolumn{1}{c}{Area} & \multicolumn{1}{c}{Thickness} & \multicolumn{1}{c}{CVI} & \multicolumn{1}{c}{Fovea} \\
\midrule
Choroidalyzer \cite{engelmann2023choroidalyzer} & 0.0381 & 9.2445 & 0.0414 & 6.9,11.0 &  & 0.0417 & 9.6963 & 0.0767 & 5.9,12.6 \\
UNet w/ ResNet18 \cite{Iakubovskii2019}  & 0.0300 & 7.1159 & 0.0348 & 12.1,11.3 &  & 0.0400 & 9.1159 & 0.0448 & 12.1,11.3 \\
UNet w/ MobileNetV3 \cite{Iakubovskii2019} & 0.0597 & 14.8460 & 0.0517 & 12.0,9.2 &  & 0.0597 & 14.8460 & 0.0517 & 12.0,9.2 \\
REACHNet \textrm{(ours)} & \textbf{0.0273} & \textbf{6.8788} & \textbf{0.0338} & \textbf{6.4,5.4} &  & \textbf{0.0350} & \textbf{8.3590} & \textbf{0.0382} & \textbf{5.0,5.5}\\
\midrule
\multicolumn{1}{r}{REACHNet --- no domain-specific augs} & 0.0306 & 7.4182 & \underline{0.0329} & 6.4,5.6 &  & 0.0431 & 10.0769 & 0.0427 & 5.2,5.7 \\
\multicolumn{1}{r}{REACHNet --- no multi-resolution} & 0.0300 & 7.7629 & 0.0383 & 10.7,8.0 &  & 0.0412 & 10.0533 & 0.0410 & \underline{5.1,5.3} \\
\multicolumn{1}{r}{REACHNet --- no self-attention} & 0.028 & \underline{6.6203} & 0.0354 & 6.5,7.0 &  & 0.0407 & 10.2387 & 0.0391 & 5.6,5.8 \\
\bottomrule
\end{tabular}}}
}
\caption{Mean absolute errors (MAE) of segmentation-derived choroid measurements in internal and external test sets. CVI, choroid vascular index. Area measured in mm$^2$, thickness in microns, CVI is dimensionless and fovea MAE is expressed in pixels for both $x$ and $y$ coordinates.}
\label{supp_tab:trainvaltest_population_overview_tab}
\end{table}

\subsection{Qualitative preference stratified by device and manufacturer}

\begin{table}[H]
\centering
\vspace*{-\baselineskip}
{\small
\scalebox{0.8}{\centerline{\begin{tabular}{@{}lccc@{}}
\toprule
Device type & REACHNet \textrm{(ours)} & Choroidalyzer \cite{engelmann2023choroidalyzer} & Ground-truth labels \\
\midrule
Topcon DRI Triton Plus ($N=12$) & \textbf{8} & 3 & 1 \\ 
Heidelberg Standard ($N=44$) & \textbf{41} & 1 & 2 \\ 
Heidelberg FLEX ($N=44$) & \textbf{36} & 2 & 6\\ 
\midrule
Total ($N=100$) & \textbf{85} & 6 & 9\\
\bottomrule
\end{tabular}}}

}
\cprotect\caption{Preference of clinical ophthalmologist for 100 examples with poorest vessel segmentation agreement between REACHNet, Choroidalyzer \cite{engelmann2023choroidalyzer} and ground-truth labels, stratified by imaging device.}
\label{tab:quality_tab} 
\end{table}

\end{document}